\documentclass[prb,amsmath,amssymb,twocolumn]{revtex4}
\usepackage{t1enc}              
\usepackage[latin2]{inputenc}   

\usepackage{graphicx}
\usepackage{dcolumn}
\usepackage{bm}

\begin{document}


\title{The effect of electron-electron interactions\\ on the conditions of surface state existence}

\author{Jarosław Kłos}
\email{klos@amu.edu.pl}

\affiliation{Surface~Physics~Division,~Faculty~of~Physics,~Adam
Mickiewicz~University,\\ ul.Umultowska 85, 61-614 Pozna\'{n},
Poland}


\begin{abstract}
Electronic surface states in one-dimensional two-band TBA model
are studied by use of the Green function method. The local density
of states (LDOS) at successive atoms in a semi-infinite chain,
even in the case of atoms distant from the surface, is found to be
clearly different from that observed in an unperturbed (infinite)
chain \cite{foo}. The surface atom occupancy is calculated
self-consistently \cite{newns}, with the effect of
electron-electron interactions taken into account. The
electron-electron interactions are shown to have a significant
impact on the conditions of surface state existence.
\end{abstract}

\maketitle

\section{Introduction}

The presence of electronic surface states has a substantial effect
on the properties of solids. Many apparently surprising features
of mesoscopic and nanoscopic systems - in which surface effects
are particularly conspicuous, the surface representing a
significant part of the whole - can be elucidated by the
conditions of existence of surface states.

Adsorption and reconstruction processes render the description of
real surfaces difficult. The properties of real systems can often
be precisely reproduced through numerical simulations, which,
however, do not provide the explanation of the mechanism of
generation of surface states and their effect on bulk states.

One of the basic models used for description of electronic
properties of solids is the tight binding approximation (TBA)
model. The pioneering studies on the conditions of electronic
surface state existence, based on a single-band model of finite
crystal, were reported by Goodwin \cite{good}. The single-band
model was then generalized by Artman \cite{art}, who introduced a
double-band model to investigate the existence of two types of
surface states: Shockley states, which are induced only by
breaking the translational symmetry of the crystal, and Tamm
states, generated as a result of introducing an additional
perturbation \cite{zak}. A breakthrough was marked by the paper by
Kalkstein and Soven \cite{sov}, in which Green's function
formalism was used for the determination of properties of surface
and bulk states in a semi-infinite crystal with perturbed surface.

The problem of multi-electron effects and their impact on surface
state existence conditions has not yet been exhaustively
discussed. Papers on chemisorption, which is a related issue, are
available, though \cite{newns, dav}. As in the case of
chemisorption, the simples way of including the impact of the
multi-electron effects on the conditions of surface state
existence consists in incorporating interaction of electron with
charge density into the Coulomb model. Introducing the Hartree
potential into the TBA model of a semi-infinite crystal amounts
(in the simplest case) to surface atom site energy
renormalization.

This study is focused on surface states in a 1D semi-infinite atom
chain being a model of ionic crystal with two atoms in the unit
cell \cite{davl}. Multi-electron effects are taken into account in
Hartree approximation only. The surface atom occupancy and site
energy values are found through self-consistent calculations using
Green's function formalism \cite{sov}.

\section{Model}

\begin{figure}
\centering
\includegraphics[width=3in]{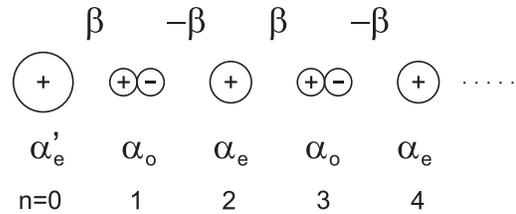}
\caption{\label{fig:f1}  The model of semi-infinite 1D crystal
with two-atom unit cell. The surface site ($n=0$) is occupied by
an adatom with site energy $\alpha'_{e}$. The alternating
resonance integral sign is a consequence of interaction between
orbitals $s$~and~$p$.}
\end{figure}

The model assumes non-zero resonance integral values for
neighboring sites only. Orbitals $s$ and $p$ are alternately
attributed to successive chain sites. Hence, resonance integrals
for successive site pairs alternate in sign, taking values $\beta$
or $-\beta$. Site energy associated with $s$ or $p$ orbital is
denoted $\alpha_{e}$ or $\alpha_{o}$, respectively (cf.
Fig.\ref{fig:f1}).

A wave function in the TBA model is assumed to be a linear
combination of atomic functions:
\begin{equation}
\left|k\right\rangle=\sum_{n}^{N}\left(c_{2n}\left|2n\right\rangle+c_{2n+1}\left|2n+1\right\rangle\right),
\end{equation}
the sum involving all the two-atom unit cells. Expressed in the
atomic function basis, the Hamiltonian of an infinite
(unperturbed) chain has the following form:
\begin{eqnarray}
\hat{H}_{0}&=&\sum_{n}\left(\alpha_{e}\left| 2n
\right\rangle\left\langle 2n \right|+\alpha_{o}\left| 2n+1
\right\rangle\left\langle 2n+1 \right|+  \right. \nonumber\\
&&\left. + \beta\left| 2n \right\rangle\left\langle 2n+1
\right|-\beta\left| 2n+1 \right\rangle\left\langle 2n
\right|\right).
\end{eqnarray}
A surface introduced into the system is regarded as a perturbation
breaking the infinite chain into two separate semi-infinite ones
\cite{sov}:
\begin{eqnarray}
\hat{H}&=&\hat{H_{0}}+\hat{V}\\
\hat{V}&=&(\alpha'_{e}-\alpha_{e})\left(\left|-1
\right\rangle\left\langle -1 \right|+\left| 0
\right\rangle\left\langle 0 \right|\right)+\nonumber\\&&
+\beta\left(\left| -1 \right\rangle\left\langle 0 \right|+\left| 0
\right\rangle\left\langle -1 \right|\right).
\end{eqnarray}
The non-zero values of elements $V\left(0,0\right)$ and
$V\left(-1,-1\right)$ allow for adsorption of atom of different
type.

Derived from the secular equation for Hamiltonian $\hat{H_{0}}$,
the expansion coefficients $c_{n}$ and the dispersion relation
read as follows:
\begin{eqnarray}
\left\{
\begin{array}{cc}
  c_{2m}=A e^{i m \theta/2} \\
  c_{2m+1}=A B e^{i m \theta/2}
\end{array},
\right.\\ X=\pm\sqrt{\tau^{2}+2-2 \cos \theta},
\end{eqnarray}
where
\begin{eqnarray}
A=\sqrt{\frac{X+\tau}{2N \xi}},&& B=\frac{2i \sin\left(
\theta/2\right)}{X+\tau},
\end{eqnarray}
and
\begin{eqnarray}
\xi= \left\{
\begin{array}{cc}
  X,  &\left|X\right|>\tau \\
  \tau,&  \left|X\right|<\tau
\end{array}.
\right.
\end{eqnarray}
Parameters $X$ and $\theta$ represent dimensionless energy and
wave vector, respectively:
\begin{eqnarray}
X=\frac{E-\bar{\alpha}}{\beta}, &&\theta=\frac{2\pi k}{N}.
\end{eqnarray}
$\bar{\alpha}$ and $\tau$ are defined as follows:
\begin{eqnarray}
\bar{\alpha}=\frac{\alpha_{e}+\alpha_{o}}{2},&&
\tau=\frac{\alpha_{e}-\alpha_{o}}{2 \beta}.
\end{eqnarray}

The Greenian matrix elements:
\begin{equation}
\hat{G_{0}}(E)=\sum_{k}\frac{\left| k \right\rangle\left\langle k
\right| }{E-E(k)}
\end{equation}
expressed in the atomic function basis, read within the energy
bands:
\begin{eqnarray}
G_{0}(2m,2n)&=&-\frac{X+\tau}{\beta}\frac{t_{<}^{n-m}}{t_{<}-t_{>}},\\
G_{0}(2m+1,2n+1)&=&-\frac{X-\tau}{\beta}\frac{t_{<}^{n-m}}{t_{<}-t_{>}},\\
G_{0}(2m,2n+1)&=&\frac{1-t_{<}}{\beta}\frac{t_{<}^{n-m}}{t_{<}-t_{>}},\\
G_{0}(2m+1,2n)&=&\frac{1-t_{<}^{-1}}{\beta}\frac{t_{<}^{n-m}}{t_{<}-t_{>}},
\end{eqnarray}
where
\begin{eqnarray}
t_{\gtrless}=Z\pm {\rm sign}(X)i\sqrt{Z^{2}-1},
\end{eqnarray}
and
\begin{equation}
Z=\frac{\tau^{2}+2-X^{2}}{2}=\cos \theta.
\end{equation}

Selected Greenian matrix elements for perturbed (semi-infinite)
crystal can be found from Dyson's equation:
\begin{equation}
\hat{G}=\hat{G_{0}}+\hat{G_{0}}\hat{V}\hat{G}.
\end{equation}
The diagonal elements read:
\begin{eqnarray}
&&G(m,m)=G_{0}(m,m)+\\
&&+\frac{G_{0}(0,m)\left(G_{0}(m,0)V(0,0)+G_{0}(m,-1)V(-1,0)\right)}{1-G_{0}(0,0)V(0,0)-G_{0}(0,-1)V(-1,0)}\nonumber
\end{eqnarray}
This allows the determination of the local density of states
(LDOS):
\begin{equation}
\rho(X,m)=-\pi \Im\left[\beta G(m,m)\right]
\end{equation} and the surface state occupancy in successive chain sites:
łańcucha:
\begin{equation}
\left\langle {\bm n}(m)\right\rangle={\rm
Res}\left[G(m,m),X_{s}\right],
\end{equation}
where $X_{s}$ is the surface state energy determined from the
condition of $G(m,m)$ zeroing.

Self-consistent renormalization of site energy at successive sites
is necessary for electron-electron interactions to be taken into
account. By defining
\begin{eqnarray}
\tau_{m}=\frac{\alpha_{m}-\bar{\alpha}}{\beta},&&\alpha_{m}=\alpha_{e}',\alpha_{o},\alpha_{e},\alpha_{o},\alpha_{e},\ldots
\end{eqnarray}
we get:
\begin{equation}
\tau'_{m}=\tau_{m}+U \left\langle {\bm n}(m)\right\rangle,
\end{equation}
where $U$ is a parameter defining interaction of electron with
charge density. Surface state localization is equivalent to state
occupancy fading inward the crystal. Therefore, the highest site
energy gradient is expected at the surface. In the first
approximation, site energy modification can concern the surface
atom only.

The surface perturbation parameter can be expressed as follows:
\begin{eqnarray}
\Delta_{e}=\frac{\alpha'_{e}-\alpha_{e}}{\beta}=\tau_{0}-\tau_{2m},&&
m=1,2,3,\ldots .
\end{eqnarray}
When interaction of electron with charge density is taken into
account:
\begin{eqnarray}
\Delta'_{e}=\tau'_{0}-\tau_{2m},&& m=1,2,3,\ldots.
\end{eqnarray}
In the case considered here ($\tau=1$), the surface state energy
is expressed by the following formula:
\begin{eqnarray}
X_{s \pm}=
  \frac{1+\Delta'^{2}_{e} \pm \sqrt{1-4\Delta'_{e}+6\Delta'^{2}_{e}+4\Delta'^{3}_{e}+\Delta'^{6}_{e}}}{2
\Delta'_{e} },
\end{eqnarray}
$X_{s+}$ and $X_{s-}$ being the solutions valid for
$\left|\Delta'_{e}+1/2\right|\nobreak>\nobreak\sqrt{5}/2$ and
$\Delta'_{e}<0$, respectively.

\section{Results}
\begin{figure}
\centering
\includegraphics[width=3in]{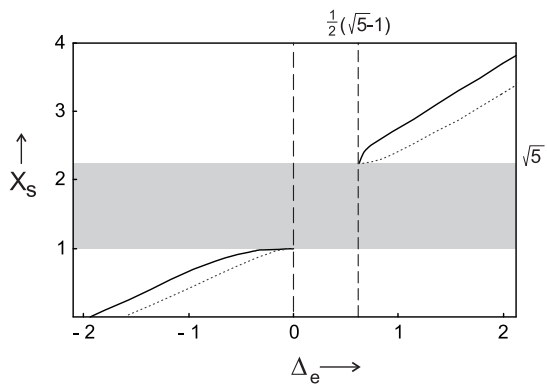}
\caption{\label{fig:f2} Surface state levels around the upper band
(gray area). The solid and dotted lines represent the state levels
found with or without the electron-electron interactions taken
into account, respectively. The dashed lines delimit the region in
which no surface states exist.}
\end{figure}

Computations were performed at $\tau\nobreak=\nobreak1$
$(\alpha_{e}-\alpha_{o}=2\beta)$. Fig.\ref{fig:f2} shows the
surface state energy, $X_{s}$, plotted versus the surface
perturbation. No Shockley states are found to exist in the model
discussed \cite{zak}, as no surface states are found in the
absence of perturbation. Tamm states, induced through modifying
the surface atom site energy, $\Delta'_{e}$, are found to emerge
from the upper energy band (associated with $\alpha_{e}$ ).

The surface states emerging from the bottom edge of the band are
induced by arbitrarily small perturbation value. For surface
states to be induced above the upper band, however, the
perturbation value must be positive and fulfill the condition
$\Delta'_{e}\nobreak>\nobreak1/2(\sqrt{5}-1)$. Perturbation values
from the interval $0\nobreak<\nobreak\Delta'_{e}\nobreak<\nobreak
1/2(\sqrt{5}-1)$ correspond to surface state non-existence. The
solid and dotted lines in Fig.\ref{fig:f2} represent surface state
levels found with or without the electron-electron interactions
taken into account, respectively. Clearly, the multi-electron
effects (in the Hartree approximation) boost the surface state
energy levels, resulting in weakened or strengthened localization
of surface states below or above the upper band, respectively.
However, the electron-electron interactions have no effect on the
interval of surface perturbation parameter values at which surface
states are found to exist.

\begin{figure*}
\centering
\includegraphics[width=6.5in]{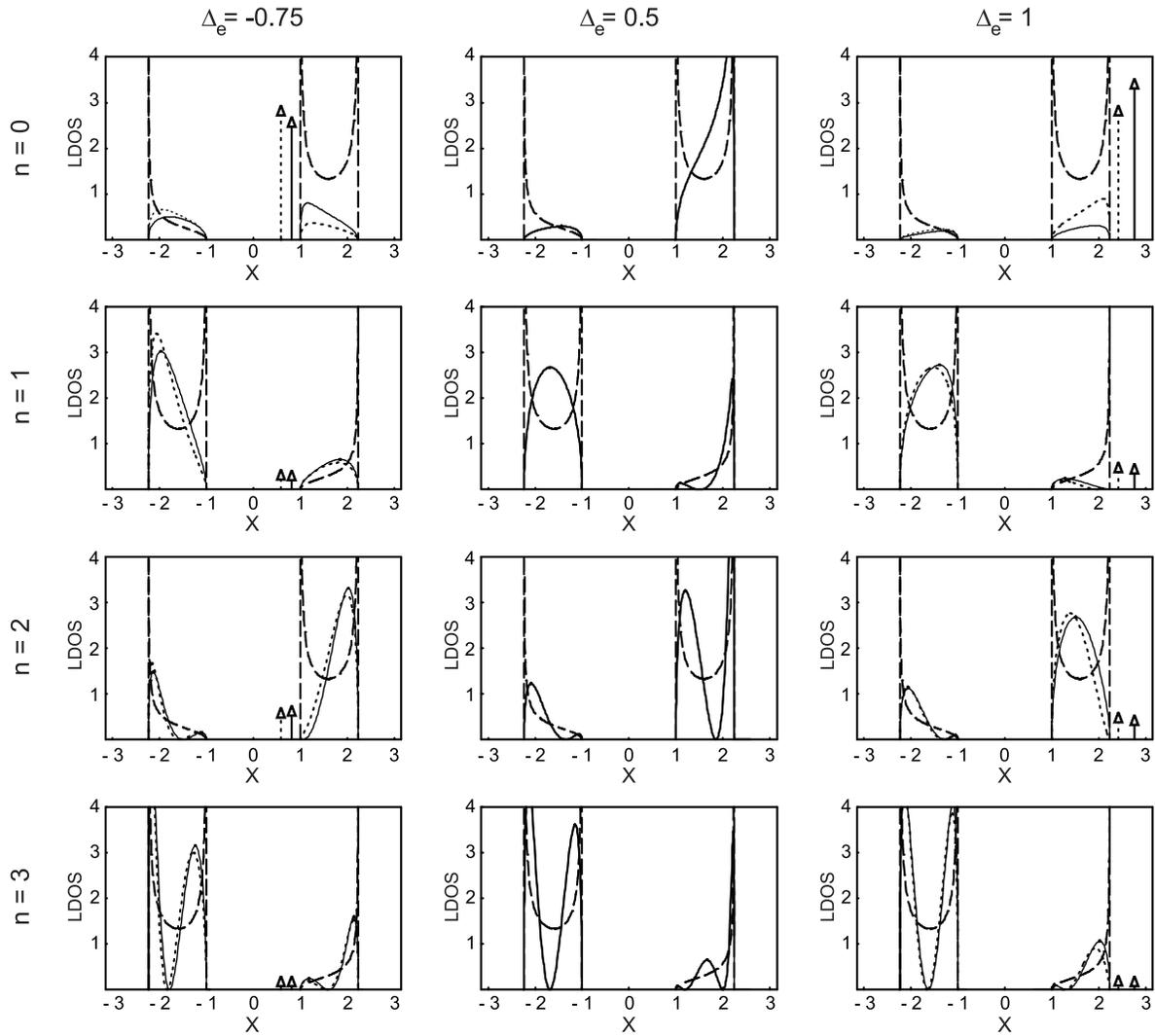}
\caption{\label{fig:f3} The local density of states,
($\pi\rho(X)$), at successive chain sites starting from the
surface ($n=0,1,2,3$). Results obtained at three different
perturbation values, $\Delta_{e}$, are grouped into columns. The
dashed line represents the LDOS for infinite chain. The dotted and
solid lines represent the LDOS found for semi-infinite chain with
electron-electron interactions neglected or taken into account,
respectively. The arrows indicate the surface state occupancy.}
\end{figure*}

The effect of the surface on the electronic states in the
considered chain is the most evident in the LDOS spectrum.
Fig.\ref{fig:f3} shows the LDOS plots obtained for the four sites
closest to the surface ($n=0 \ldots 3$). Three different
perturbation values are assumed, corresponding to surface state
appearing below the upper band
($\Delta_{e}\nobreak=\nobreak-0.75$), not induced at all
($\Delta_{e}\nobreak=\nobreak0.5$), and induced above the upper
band ($\Delta_{e}\nobreak=\nobreak1$). The solid and dotted lines
represent the LDOS calculated with multi-electron effects taken
into account or neglected, respectively. As a result of including
the multi-electron effects, the LDOS in the upper band is
increased; at the same time, the occupancy of the surface state
below the band decreases, and the state moves towards the band
edge. An opposite effect is found to occur for the surface state
above the upper band: the LDOS in the band is found to decrease,
while the surface state occupancy increases and the state moves
inwards the gap. Comparing relative occupancy changes at
successive sites, one notes stronger localization in states closer
to the band edge. Because of band asymmetry, the surface state
occupancy should be compared between sites either even or odd.
With multi-electron effects taken into account, the occupancy
ratio of site ($n=0$) (the surface atom) to site $n=2$ is found to
increase or decrease for states above or below the upper band,
respectively. Note that even in the absence of surface states the
presence of the surface still affects the LDOS spectrum. The van
Hove singularities at the band edges are eliminated, and LDOS
minima appear inside the bands. Even for sites very distant from
the surface, the LDOS differs significantly from that in an
infinite (unperturbed) chain \cite{foo}.


\section{Conclusions}

Focused on the effect of electron-electron interactions on the
conditions of surface state existence (in the Hartree
approximation), the above-presented study shows that the
localization of surface states generated above or below the upper
energy band can be increased or decreased, respectively, by the
electron-electron interactions in the considered model. However,
the interval of surface perturbation parameter values
corresponding to surface state existence is found to remain
unaffected by taking these interactions into account in the
calculations.

\end{document}